\documentclass[aps,prl,twocolumn,superscriptaddress,showpacs]{revtex4}

\usepackage{amssymb}
\usepackage{amsmath}
\usepackage{graphicx}
\usepackage{natbib}
\usepackage{nicefrac}
\usepackage{multirow}
\usepackage{comment}

\begin{document}

\title{Landau-Zener-St\"{u}ckelberg Interferometry of a Single Electron Charge Qubit}

\author{J. Stehlik}
\affiliation{Department of Physics, Princeton University, Princeton, New Jersey 08544, USA}
\author{ Y. Dovzhenko}
\affiliation{Department of Physics, Princeton University, Princeton, New Jersey 08544, USA}
\author{J. R. Petta}
\affiliation{Department of Physics, Princeton University, Princeton, New Jersey 08544, USA}
\affiliation{Princeton Institute for the Science and Technology of Materials (PRISM), Princeton University, Princeton,
New Jersey 08544, USA}
\author{J. R. Johansson}
\affiliation{Advanced Science Institute, RIKEN, Wako-shi, Saitama 351-0198, Japan}
\author{F. Nori}
\affiliation{Advanced Science Institute, RIKEN, Wako-shi, Saitama 351-0198, Japan}
\affiliation{Physics Department, University of Michigan, Ann Arbor, Michigan 48109, USA}
\author{H. Lu}
\affiliation{Materials Department, University of California at Santa Barbara, Santa Barbara, California 93106, USA}
\author{ A. C. Gossard}
\affiliation{Materials Department, University of California at Santa Barbara, Santa Barbara, California 93106, USA}

\begin{abstract}
We perform Landau-Zener-St\"{u}ckelberg interferometry on a single electron GaAs charge qubit by repeatedly driving the system through an avoided crossing. We observe coherent destruction of tunneling, where periodic driving with specific amplitudes inhibits current flow. We probe the quantum dot occupation using a charge detector, observing oscillations in the qubit population resulting from the microwave driving. At a frequency of 9 GHz we observe excitation processes driven by the absorption of up to 17 photons. Simulations of the qubit occupancy are in good agreement with the experimental data.
\end{abstract}


\maketitle

Semiconductor quantum dots are fruitful systems for exploring phenomena arising from quantum interference effects \cite{chargequbit1,Fujisawa20041046,PettaSeminal,Koppens,fol,GefenPRL}. Landau-Zener-St\"{u}ckelberg (LZS) interferometry has recently emerged as a novel way to study quantum coherence in solid state systems. LZS theory was initially described in the context of atomic collisions and relies on having an effective two-level system with an avoided crossing in the energy level spectrum \cite{landau,zener,stuck1,Majorana,Shimshoni}. Repeated sweeps through the avoided crossing result in successive Landau-Zener transitions, allowing control of the final state probability. While the theory was initially applied to atomic collisions, recent advances in the fabrication of solid state quantum devices have made it experimentally accessible in a wide variety of systems, ranging from superconducting qubits \cite{ShevchenkoReview} to nitrogen vacancy centers in diamond \cite{NVlzs,fuchsNVZ}. In superconducting qubits, LZS interferometry has been used with great success to determine the energy level diagram and to measure qubit coherence times  \cite{ShevchenkoReview,Oliver,LZS_cooper}. In spin qubits, LZS interferometry has been harnessed to drive coherent singlet-triplet transitions resulting in spin rotations that are much faster than those obtained using conventional electron spin resonance \cite{LZ_petta,Sachrajda}.

In this Rapid Communication we perform LZS interferometry on a single electron GaAs double quantum dot (DQD) charge qubit. The sample geometry is illustrated in the scanning electron microscope (SEM) image shown in Fig.\ 1(a). Ti/Au gate electrodes are fabricated on top of a GaAs/AlGaAs heterostructure that is grown using molecular beam epitaxy. The gate electrodes selectively deplete regions of the two-dimensional electron gas located 110 nm below the surface of the wafer, forming a DQD containing a single electron. In this experiment, a third dot is used as a charge detector, which allows non-invasive measurements of the charge state occupancy \cite{QPC}. A fixed 100 mT field is applied perpendicular to the plane of the sample. Despite their simplicity, charge qubits are of great experimental importance as they allow for direct quantum control through electric fields, with coherent control rates dictated by tunnel couplings that can easily approach 10 GHz. They also serve as building blocks for more complex quantum systems, such as spin qubits \cite{Loss}.

\begin{figure}[t]
\begin{center}
\includegraphics[width=\columnwidth]{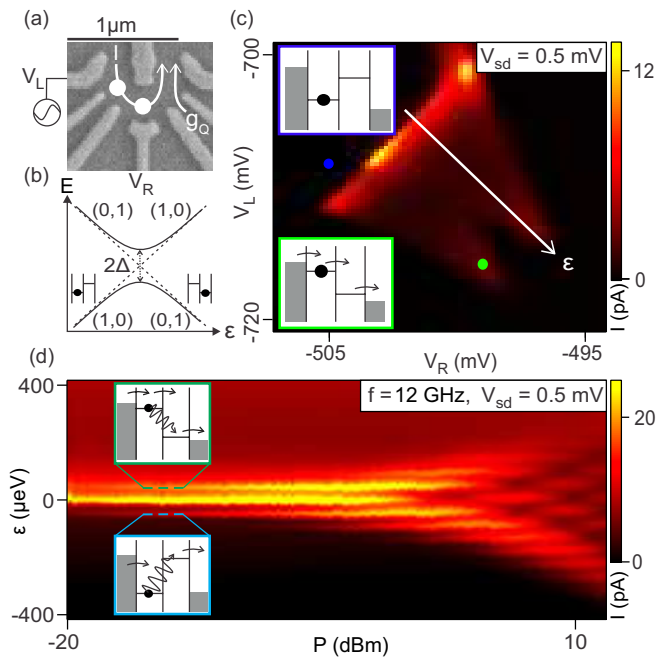}
\caption{\label{fig1} (Color online) (a) SEM image of a device similar to the one measured.  (b) Energy spectrum plotted as a function of detuning. (c) Current through the DQD as function of $V_{\rm L}$ and $V_{\rm R}$. The detuning axis used for finite-bias measurements is plotted in white. Insets: Energy level diagrams for two different locations in the charge-stability diagram. (d) Current through the DQD as function of detuning and applied microwave power at $f$ = 12 GHz. The application of microwaves to gate $V_{\rm L}$ drives transitions from the ground state to the excited state on resonance.  Insets illustrate tunneling processes that are driven by the absorption of one photon for positive and negative detuning.}
\end{center}	
\vspace{-0.6cm}
\end{figure}

We focus on the one electron regime, where the DQD contains a single charge. We label the charge states ($N_{\rm L}$, \nolinebreak  $N_{\rm R}$), where $N_{\rm L}$ ($N_{\rm R}$) is the number of electrons in the left (right) dot. In the charge basis, the single electron can either occupy the left dot or the right dot, corresponding to the (1,0) or (0,1) charge state, leading to the Hamiltonian:
\begin{equation}
H_0 = \frac{\epsilon}{2} \sigma_{\rm z} + \Delta \sigma_{\rm x}.
\end{equation}
Here the detuning $\epsilon$ sets the energy difference between the two dots. The qubit level splitting is given by $\Omega = \sqrt{\epsilon^2+4\Delta^2}$, where the tunnel coupling $\Delta$ results in a minimum splitting of 2$\Delta$ at $\epsilon$ = 0. The resulting energy level diagram is shown in Fig.\ 1(b).

Adding a sinusoidal driving term to the Hamiltonian,
\begin{equation}
H_{\rm t} = \frac{e V_{\rm ac}}{2} \sigma_{\rm z} \sin(\omega t),
\end{equation}
turns the two-level system into a solid-state equivalent of the optical Mach-Zehnder interferometer \cite{machzenner}.  In the limit that $e V_{\rm ac}$ is large compared to $\Delta$, the qubit traverses the avoided crossing twice at approximately constant velocity during each cycle of the driving field. The probability that an electron initially in the ground state will transition to the excited state during one such traversal is given by the Landau-Zener formula \cite{ShevchenkoReview},
\begin{equation}
P_{\rm LZ} = \exp\left(-2 \pi \frac{\Delta^2}{\hbar v }\right).
\end{equation}
Here $v$ = $\mathrm{d} E / \mathrm{d}t $ is the level velocity, where $E$ is the energy difference of the uncoupled levels. Away from the avoided crossing, the excited and ground states evolve independently and acquire a St\"{u}ckelberg phase, which is a function of the time spent between the crossings and $V_{\rm ac}$. The two Landau-Zener transitions are the effective beam-splitters of the Mach-Zehnder interferometer.

Mach-Zehnder interferometry was previously demonstrated in superconducting flux qubits \cite{Oliver} and singlet-triplet qubits \cite{LZ_petta}. While in a superconducting flux qubit $ \Delta $ is fixed once the sample is made, in our system it can be tuned \emph{in situ} by adjusting gate voltages. Thus a gate-defined charge qubit can access both weak and strong interdot tunnel couplings with a single device. However, the extra tunability comes with a price, as electrical driving can also modulate the $\sigma_{x}$ term in the Hamiltonian, complicating the charge dynamics.

We probe LZS interferometry by utilizing two different measurements.  We first examine photon-assisted transport by applying a source-drain bias and measuring the current through the DQD in the presence of microwave driving
\cite{RevModPhys.75.1,Aguado,KouwenhovenPat,KouwenhovenPat2}.  In this measurement the microwave driving can transfer population from the ground state to the excited state when the microwave photon energy matches the energy splitting of the charge qubit levels, resulting in charge pumping through the sample.  With this configuration the total current through the DQD is sensitive to the microwave coupling as well as to the overall coupling of the DQD to the source and drain electrodes. We also probe the DQD occupation using the charge detector, which directly measures the occupation of the left quantum dot, $P_{(1,0)}$. Charge sensing is performed without a source-drain bias, thereby probing the DQD in a manner which is relatively insensitive to the coupling of the DQD to the leads.

Transport through the DQD is measured by applying a source-drain bias across the device, $V_{\rm sd} = 0.5  \:\mathrm{mV}$.  Due to the discrete energy levels of the quantum dots, current can only flow through the device when the energy levels of the two dots are within the energy window set by $eV_{\rm sd}$ and the sign of the detuning matches the sign of the source-drain bias. As a result, a non-zero current is only observed in finite-bias triangles located near triple points in the charge-stability diagram, as shown in Fig.\nolinebreak\ 1(c) \cite{RevModPhys.75.1}. By periodically driving the qubit, LZS interference can result in transitions from the ground state to the excited state, effectively driving an uphill tunneling process. In the limit of fast driving ($e V_{\rm ac}  \hbar \omega \gg \Delta^2$) constructive interference of the St\"{u}ckelberg phase can only occur if the qubit splitting $\Omega$ = $n\hbar \omega$, with $n$ some integer. This can be readily identified as a \emph{n}-photon process. We plot the current through the DQD as a function of detuning and applied microwave power at a fixed frequency $f$ = 12 GHz in Fig.\ 1(d).  The current oscillates as a function of detuning due to multi-photon absorption and as a function of power due to changes in the St\"{u}ckelberg phase.
\begin{figure*}
\begin{center}
		\includegraphics[width=2\columnwidth]{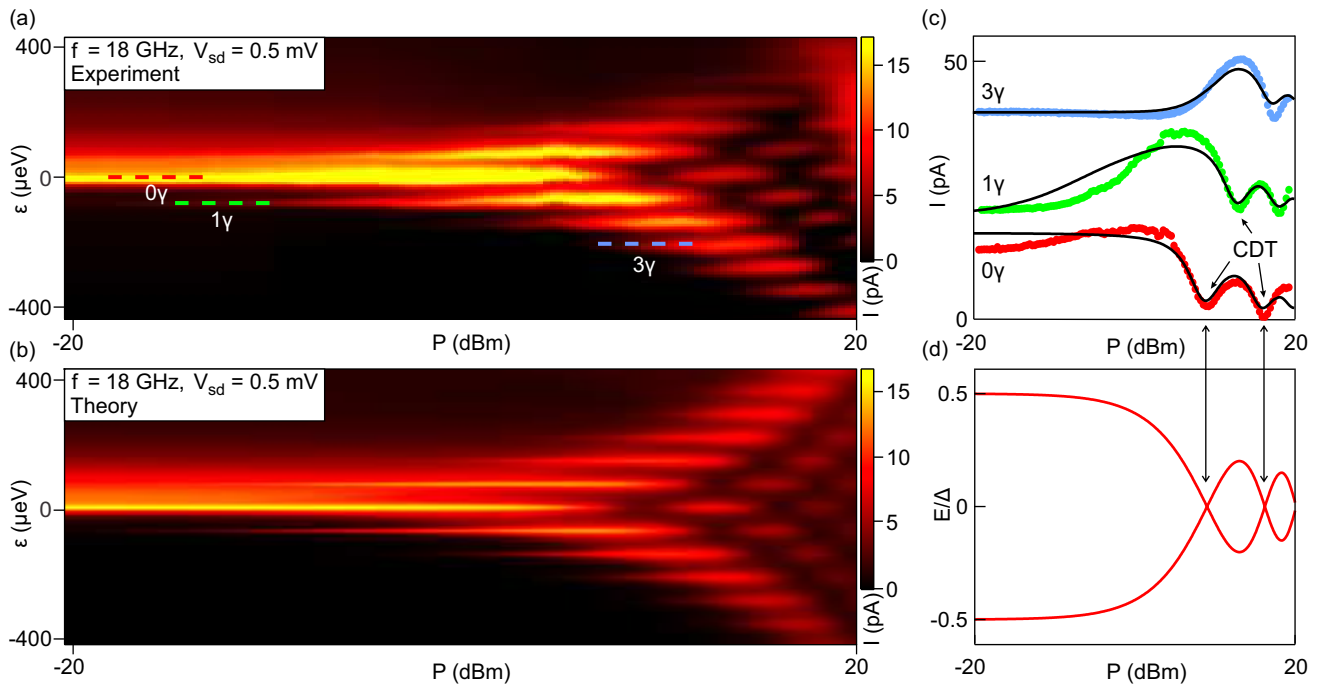}
\caption{\label{fig2} (Color online) (a) Current through the double dot measured as function of detuning and applied microwave power at $f = 18 \:\mathrm{GHz}$ and $V_{\rm sd} = 0.5 \: \mathrm{mV}$. (b) Current obtained from numerical simulations. (c) Current extracted from the data in (a) along the 0, 1, and 3 photon peaks (successive traces are offset by 20 pA for clarity). Regions of suppressed current are due to the coherent destruction of tunneling. (d) Predictions of the two lowest quasi-energies arising from the Floquet formalism \cite{PhysRevB.67.165301}. Quasi-energy crossings are indicative of vanishing wavefunction overlap, leading to the coherent destruction of tunneling.}
\end{center}	
\vspace{-0.5cm}
\end{figure*}

We further investigate the current through the DQD in Fig.\ 2(a), where the microwave frequency is set to $f$ = 18 GHz. The interference pattern can be calculated in the LZS framework and arises from coherent interference due to the St\"{u}ckelberg phase. Similar interference patterns have been observed in many-electron DQDs \cite{Oosterkamp1PAT,Fujisawa1PAT,OnoPhysicaE}. We model the data using a simple three-level system consisting of the (0,0), (0,1) and (1,0) charge states. We allow electrons to incoherently tunnel from the (0,1) state to the (0,0) state, and from the (0,0) state to the (1,0) state, with rates $\Gamma_{\rm R} = \Gamma_{\rm L} = 320 \: \mathrm{MHz}$, simulating the usual transport cycle in DQD devices.  The (1,0) and (0,1) states of the qubit are tunnel-coupled with $\Delta/h = 400 \: \mathrm{MHz}$.  In addition the higher energy level is allowed to inelastically tunnel into the lower level.  The rate for this process and its detuning dependence are extracted from measurements of the current as a function of detuning in the limit of no microwave driving \cite{Brandes,FujisawaEm, som1}. The inelastic tunneling rate at zero-drive ranges from 270 MHz near zero-detuning to 80 MHz at large detunings. We assume the dephasing rate is an increasing function of driving power, as driving with a large amplitude results in higher electron temperatures. The dephasing time in the model ranges from $600\:\mathrm{ps}$ at zero-drive to $120 \: \mathrm{ps}$ at $P=20 \:\mathrm{dBm}$ {\cite{som1}. The resulting plot of the simulated current is shown in Fig.\ 2(b).

In both Fig.\ 1(d) and Fig.\ 2(a), a Bessel function modulation of the current is observed as a function of microwave power. The power dependence can be more clearly seen in Fig.\ 2(c), where the dot current is plotted as a function of microwave power at detunings corresponding to the 0, 1, and 3 photon resonant peaks ($0 \gamma,\: 1 \gamma,\: 3 \gamma$). At specific microwave driving powers, the current is strongly suppressed despite a finite source-drain bias. The zeros in the current are due to the coherent destruction of tunneling (CDT) \cite{PhysRevLett.67.516,Grifoni}. In the LZS framework, CDT occurs at conditions for which the St\"{u}ckelberg phase results in perfect destructive interference. However, in this case, it is more intuitive to consider the tunneling process using Floquet theory, which is an analogue of Bloch's theory for time-periodic systems as opposed to space-periodic ones. The long-term evolution of a periodically-driven system is described by quasi-energies, which naturally allow for multiple-photon resonances. The quasi-energy spectrum can exhibit exact crossings as shown in Fig.\ 2(d), which plots the lowest two quasi-energies to leading order in perturbation theory \cite{PhysRevB.67.165301}. The crossings (with no gap) imply that the effective interdot tunnel rate goes to zero, resulting in current suppression \cite{PhysRevLett.67.516}.
\begin{figure*}
\begin{center}
		\includegraphics[width=2\columnwidth]{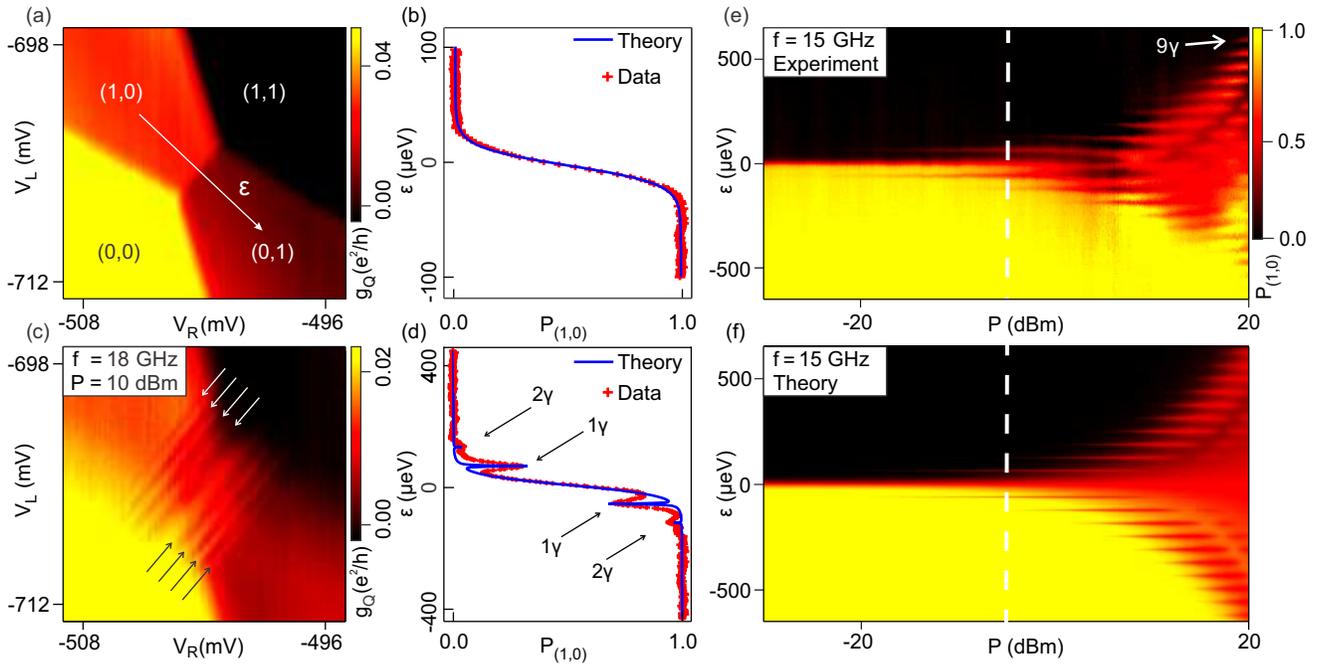}
\caption{\label{fig3} (Color online) (a) Charge detector conductance as a function of gate voltages $V_{\rm L}$ and $V_{\rm R}$. The detuning axis traverses the (1,0)--(0,1) charge transition. (b) Left dot occupation, $P_{(1,0)}$, measured as a function of detuning in the absence of microwave driving. A fit to the data yields an electron temperature $T_{\rm e} = 110 \: \mathrm{mK}$. (c) Charge detector conductance as a function of $V_{\rm L}$ and $V_{\rm R}$ with the microwave driving applied, showing resonant peaks due to photon absorption. (d) $P_{(1,0)}$ measured as a function of detuning with a 15 GHz microwave driving frequency. Resonant excitation drives transitions from the ground state to the excited state, modifying the qubit population. (e) $P_{(1,0)}$ oscillates as a function of detuning and applied microwave power at 15 GHz. We observe tunneling processes that are driven by the absorption of 9 photons. A background subtraction is performed \cite{som1}. (f) $P_{(1,0)}$ obtained from numerical simulations plotted on the same color-scale as the experimental data \cite{som1}.}
\end{center}
\vspace{-0.6cm}
\end{figure*}

While providing good agreement between theoretical predictions and measurements, transport studies are burdened by the complication of electron tunneling between the leads and the DQD. Charge sensing allows us to directly measure the charge occupancy of the DQD in the presence of periodic driving. Figure 3(a) shows the charge detector conductance, $g_{\rm Q}$, measured as function of $V_{\rm L}$ and $V_{\rm R}$ showing the expected DQD charge-stability diagram. A measurement of $g_{\rm Q}$ as a function of detuning is shown in Fig.\ 3(b). The width of the transition is set by the interdot tunnel coupling and the electron temperature $T_{\rm e}$ \cite{DiCarlo}. For this device tuning $\Delta \ll k_{\rm B} T_{\rm e}$ and we extract $T_{\rm e} = 110\: \mathrm{mK}$.

Applying microwaves to the system drives transitions from the ground to the excited state when the energy level splitting is an integer multiple of the photon frequency; the resonance requirement for a LZS transition in the fast driving regime ($e V_{\rm ac} \hbar \omega \gg \Delta^2$). As a result, the measured charge detector response, shown in Fig.\ 3(c), exhibits deviations from the ground state occupation measured in Fig.\ 3(a).  Clear $1\gamma$, $2\gamma$, and $3\gamma$ transitions are observed. The LZS interference pattern is measured as a function of detuning and microwave power in Fig.\ 3(e) for $f$ = 15 GHz. A clear interference pattern is observed that exhibits many of the features found in the dc current measurements. Microwave driving affects the response of the charge detector; therefore a background subtraction procedure was implemented that normalizes the charge detector response to the values obtained at large positive and negative detunings \cite{som1}. A similar data set is shown in Fig.\ 4(a) for a smaller driving frequency of $f$ = 9 GHz.  Here we observe charge transfer processes driven by the absorption of up to 17 photons.  The asymmetry in Fig.\ 4(a) at $ P > 0 \:\mathrm{dBm}$ is due to the presence of another energy level that is accessed at large driving amplitudes.

\begin{figure}
\begin{center}
		\includegraphics[width=\columnwidth]{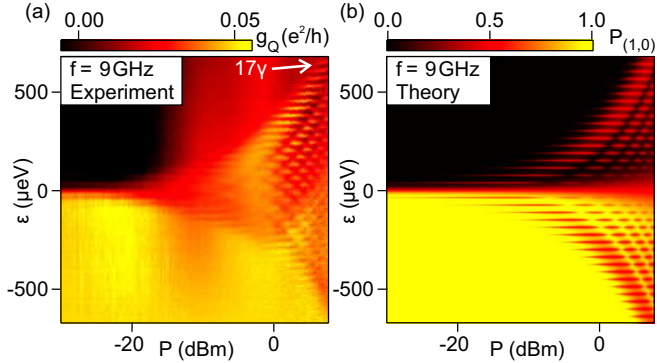}
\caption{\label{fig1} (Color online) (a) Charge detector conductance measured as a function of detuning and microwave power for $f$ = 9 GHz. With this lower driving frequency we observe up to 17 photon transitions. (b) Left dot occupation, $P_{(1,0)}$, generated from numerical simulations \cite{som1}. Simulations are in qualitative agreement with the measured data.}
\end{center}	
\vspace{-0.6cm}
\end{figure}

To establish the accuracy of the model Hamiltonian we simulated the qubit occupation in the presence of driving, relaxation, and dephasing. As in previous simulations, we assume a detuning and power-dependent relaxation rate. For this data set, the power-dependent dephasing rate is lower because the DQD states are more weakly coupled to the leads \cite{som1}. Simulations are performed by numerically calculating the steady state of the density matrix and are shown in Figs.\ 3(f) and 4(b). We obtain good agreement, indicating that the observed behavior is indeed due to LZS interferometry. Slight deviations from the theoretical model are due to direct capacitive coupling between the charge detector and the driving gate, as well as the presence of $\sigma_{\rm x}$ driving. Furthermore, the simulations reveal that the LZS oscillations should not be visible for $T_2$ less than $\sim$ 250 ps, thus establishing a lower bound on the decoherence time, consistent with what was measured using other techniques \cite{Karl,yuliya}.

In summary, we studied the dynamics of a single electron charge qubit in the presence of strong driving.  With a finite source-drain bias we observed the coherent destruction of tunneling.  Utilizing the charge detector in the zero bias regime allowed us to directly observe oscillations of the qubit occupancy predicted by the LZS theory.

Acknowledgements: Research at Princeton was supported by the Sloan and Packard Foundations, DARPA QuEST award HR0011-09-1-0007 and the National Science Foundation through the Princeton Center for Complex Materials, DMR-0819860 and CAREER award, DMR-0846341. Work at UCSB was supported by DARPA award N66001-09-1-2020 and the UCSB NSF MRSEC, DMR-1121053. J.\ R.\ Johansson was supported by a JSPS Fellowship, and F. Nori was partially supported by the ARO, JSPS-RFBR contract No.\ 12-02-92100, Grant-in-Aid for Scientific Research (S), MEXT Kakenhi on Quantum Cybernetics, and the JSPS via its FIRST program.

\end{document}